\documentclass[useAMS,usenatbib, usegraphicx]{mn2e}

\usepackage{epsfig}

\usepackage{graphicx}

\usepackage{amsmath}

\usepackage{amssymb}

\usepackage{natbib}

\newcommand{\infinity}{{\infty}}

\newcommand{\apj}{ApJ}

\newcommand{\mnras}{MNRAS}

\newcommand{\apjs}{ApJS}

\newcommand{\nat}{Nat}

\begin{document}

\title[Light-Cone Effects on Massive Galaxies]{Light-Cone Distortion of the
Clustering and Abundance of Massive Galaxies at High Redshifts}

\author[Mu{\~n}oz \& Loeb]{Joseph A.\ Mu{\~n}oz
\thanks{E-mail:jamunoz@cfa.harvard.edu} and Abraham Loeb
\thanks{E-mail:aloeb@cfa.harvard.edu}\\Harvard-Smithsonian Center for
Astrophysics, 60 Garden St., MS 10, Cambridge, MA 02138, USA\\}


\maketitle

\begin{abstract}

Observational surveys of galaxies are not trivially related to single-epoch snapshots from computer simulations.  Observationally, an increase in the distance along the line-of-sight corresponds to an earlier cosmic time at which the properties of the surveyed galaxy population may change.  The effect of observing a survey volume along the light-cone must be considered in the regime where the mass function of galaxies varies exponentially with redshift.  This occurs when the halos under consideration are rare, that is either when they are very massive or observed at high-redshift.  While the effect of the light-cone is negligible for narrow-band surveys of Ly$\alpha$ emitters, it can be significant for drop-out surveys of Lyman-break galaxies (LBGs) where the selection functions of the photometric bands are broad.  Since there are exponentially more halos at the low-redshift end of the survey, the low-redshift tail of the selection function contains a disproportionate fraction of the galaxies observed in the survey.  This leads to a redshift probability distribution (RPD) for the dropout LBGs with a mean less than that of the photometric selection function (PHSF) by an amount of order the standard deviation of the PHSF.  The inferred mass function of galaxies is then shallower than the true mass function at a single redshift with the abundance at the high-mass end being twice or more as large as expected.  Moreover, the statistical moments of the count of galaxies calculated ignoring the light-cone effect, deviate from the actual values.

\end{abstract}


\section{Introduction}

It has become common practice to analyse and interpret the observed
abundance and distribution of high-redshift galaxies by approximating a
limited survey volume to a single-epoch snapshot of the Universe
\citep{Ellis07}. The observed data is then compared to theoretical predictions
which were calculated for an idealized snapshot of this nature.  However,
in actual observations, an increase in the distance along the line-of-sight
corresponds to an earlier cosmic time at which the properties of the
surveyed galaxy population may change.

The ``snapshot approximation'' is adequate for galaxy surveys at low
redshifts, when galaxy halos are common and their mass function is not
evolving rapidly with cosmic time. At these low redshifts, a relatively
small region of space spanning a narrow redshift range can still be sufficiently large to contain an adequate sample of these
abundant objects. However, the validity of the approximation should be
carefully examined at high redshifts when massive galaxies are rare and
their abundance varies exponentially with redshift.

To illustrate the situation at high redshifts, let us consider two regions
of the same shape centered at different redshifts and containing the same
number of galaxy halos of a particular mass. The region centered at the
higher redshift will span a larger range in redshift for two reasons.
First, since halos of a given mass are rarer at a higher redshift, the
higher redshift region must have a larger comoving size than the one at
smaller redshift for each to contain the same number of halos.  Second, the
same comoving distance corresponds to a larger redshift interval at higher
redshift than at lower redshift (i.e. $dz = [H(z)/c] d\chi$, where $\chi$
is the comoving length and the Hubble parameter, $H(z)$, is an increasing
function of $z$).  The difference between snapshot analysis (on a
space-like hypersurface) and that along the light-cone is becoming
increasingly relevant with purported discoveries of very massive galaxies
near $z = 6$ \citep{Mobasher05} and as new surveys probe redshifts up to
$z=10$ \citep{Bouwens06,Iye06,Stark07}.  Even at high-redshifts, narrow-band
surveys of Lyman-alpha emitters (LAEs) span such a small range of redshifts
that they are unaffected by the exponential change in the mass function of
halos with redshift.  However, the photometric selection functions (PHSFs)
of the bands used in drop-out surveys of Lyman-break galaxies (LBGs) can be
fairly broad in redshift space \citep{BI06}.

In this paper, we examine the significance of light-cone distortions on the
inferred abundance and clustering properties of high-redshift galaxies in
dropout surveys of LBGs.  First, we describe the PHSFs used in dropout
surveys in \S \ref{selection}.  Subsequently, we derive analytic formulae
for the first and second statistical moments of the count of halos in a
given survey volume (\S \ref{moments}) and consider the two-point
correlation function of halos on the light-cone (\S \ref{correlation}).  In
\S \ref{model}, we review a simple model for high-redshift star forming
galaxies by \citet{SLE07}, that gives the luminosity
of LBGs and LAEs contained in a halo of a given mass.  We then use this
model to calculate the quantitative difference between our light-cone
formalism and the standard snapshot approach for various survey volumes (\S
\ref{results}), exploring the dependence on cosmological parameters.
Finally, we discuss the significance of our results in \S \ref{discussion}.

Unless otherwise stated, we assume a flat, $\Lambda$CDM cosmology with cosmological parameters $\left( \Omega_m, \Omega_{\Lambda}, \Omega_b, h, \sigma_8, \alpha, r\right) = \left( 0.268, 0.732, 0.042, 0.704, 0.776, 0.947, 0.000 \right)$ \citep{Spergel07}.  All
distance scales are comoving.

\section{The Dropout Selection Function}\label{selection}

Dropout surveys at high redshifts ($z\ga 6$) select LBGs by measuring a
drop in flux shortward of the Ly$\alpha$ wavelength (due to absorption by
intergalactic hydrogen).  This requires comparing the observed flux in different photometric bands.
The filter for each band is described by a profile that indicates how much
light is transmitted at each wavelength.  This transmission profile
provides a probability distribution for the wavelength of a given photon
that has passed through the filter.  Since the edge of the Ly$\alpha$
absorption trough appears at a wavelength corresponding to the redshift of
the observed galaxy, the filter profile can be expressed in redshift space
as the photometric selection function (PHSF) for a given photometric band,
which gives the distribution of the surveyed galaxies over redshift
\citep{BI06}.  The volume of the survey is an integral over this function
\citep{Steidel99}.  In this paper we focus on dropout selections in the i-,
z-, and J-bands corresponding to the standard HST filters F775W, F850LP,
and F110W, respectively.  The PHSFs for each band depend on the specific
selection criteria chosen, but are roughly approximated by Gaussians
(Bouwens 2007, personal communication).  We take the mean redshifts of the
i-, z-, and J-band PHSFs to be $\mu_z = 6.5$, $7.4$, and $10$,
respectively, and their standard deviations to be $\sigma_z = 0.5$, $0.5$,
and $1.0$.  We also ignore, for simplicity, possible interlopers at lower
redshifts whose spectra mimic those of LBGs at higher redshift as a result
of dust attenuation.

Due to the evolution of the mass function of galaxy halos within the survey
volume, the probability distribution for the redshift (RPD) of a galaxy in
the survey is not the same as the PHSF.  Even though the contribution from
galaxies in the Gaussian tail of the PHSFs is exponentially suppressed, the
density of the rare halos that contain the observed galaxies is
theoretically expected to be exponentially higher toward the low-redshift
end of the survey.  The volume per redshift interval also changes within
the survey since the area of the survey perpendicular to the line-of-sight
and the comoving distance per redshift interval along the line-of-sight are
both redshift dependent, but this is a small correction.

\begin{figure}
\begin{center}
\includegraphics[width=\columnwidth]{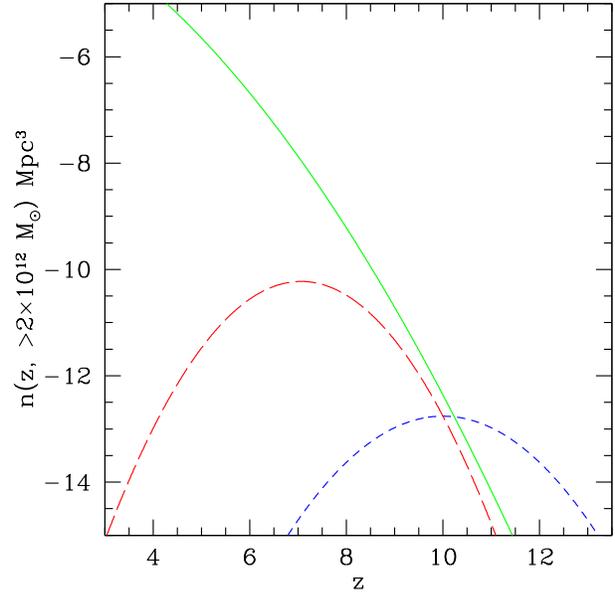}
\caption{\label{Nplot} An example of how the PHSF (dashed) is multiplied by
the mass function (solid) to yield the RPD (long-dashed).  The PHSF shown
is for the J-band, and the Gaussian is normalized to the mass function at
$z=10$ for easy viewing.  The mass function and resulting RPD were
calculated for galaxy halos with $M_{halo} > 2 \times 10^{12}\,M_{\odot}$.}
\end{center}
\end{figure}

Ignoring the variation of the survey volume per redshift interval, the RPD
for LBGs with luminosity at a wavelength of $1500\,\mbox{\AA}$ that is
greater than $L_{1500}$ within the volume observed in a given dropout band
is given by,
\begin{equation}\label{Pzg}
P_g(z) = \frac{n(z,>L_{1500}(M_{halo})))\,e^{-\frac{(z-\mu_z)^2}{2\,\sigma_z^2}}}{\int_0^\infinity n(z,>L_{1500}(M_{halo}))\,e^{-\frac{(z-\mu_z)^2}{2\,\sigma_z^2}}\,dz},
\end{equation}
where $\mu_z$ and $\sigma_z$ are the mean and standard deviation of the
given band, $n(z,M)$ is the mass function of halos, and $L_{1500}(M_{halo})$ is a
relation for the luminosity of an LBG contained in a halo of mass $M_{halo}$, which we
describe in \S \ref{model}.  Figure~\ref{Nplot} shows how the PHSF is
multiplied by the mass function to generate the real RPD for J-dropouts.
In \S \ref{results}, we calculate the moments of this true distribution.

\section{Moments of a Count of Objects}\label{moments}

In this section we derive and discuss the formulae for the statistical
moments of a count of halos above a given minimum mass, $M_{min}$, in a
given survey volume taking into account the variations of the mass and
correlation functions along the light-cone.  We limit the discussion to the
first and second moments, $\left<N\right>$ and
$\left<(N-\left<N\right>)^2\right>$.

We generally follow \citet{Peebles80} in calculating moments of a count of
objects in a box but augment the derivation by allowing the mass and
correlation functions to vary with redshift.  We begin by dividing the box
into infintessimal units of size $dV$.  The $i$--$th$ unit in the box
contains $N_i$ objects, so that the total number of objects in the box is
$N=\sum_i N_i$.  If every part of the box is viewed at the same cosmic
time, then the average number of objects in each unit is obtained through the average number
density of objects in the box: $\left<N_i\right> = n\,dV$.
However, if the box is viewed at some distance from the observer, then each
unit sits at a particular redshift, and the average number of objects in
units at redshift $z$ is related to the average number density of
objects in the box at that redshift, $\left<N_{i,z}\right> =
n(z)\,dV$.  We assume that $dV$ is small enough so that $N_i=\left\{1,0\right\}$ and
the statistics of galaxies within each unit is Poisson distributed,
i.e. $\left<N_i^k\right> = \left<N_i\right>$ at each redshift for every $k$
in $\mathbb{Z}$.

We can now calculate the moments of a count of objects, N, in the box.  The
first moment is given by
\begin{eqnarray}\label{Nave}
\left<N\right> & = & \left<\sum_i N_i\right> \nonumber \\ & = & \int n(z)\,dV,
\end{eqnarray}
where the comoving volume element $dV$ depends on the survey geometry.  In the specific case of halos above a given mass threshold $M_{min}$, $n$ is the mass function of halos above $M_{min}$.  Equation~(\ref{Nave}) is precisely what one would expect from simply integrating the mass
function over the survey volume as in \citet{NNB06}.

The second moment is
\begin{eqnarray}\label{secondmoment1}
\left<N^2\right> & = & \left<{\left(\sum_i N_i\right)}^2\right> \nonumber \\ & = & \sum_i \left<N_i^2\right> + \sum_i \sum_j \left<N_i\,N_j\right>.
\end{eqnarray}
If there is an object in each of the disjoint units $i$ and $j$, the product $N_i\,N_j$ is equal to unity.  Otherwise, the product is equal to zero.  The probability of both units containing an object is
\begin{equation}\label{jointprob}
dP = n_i\,n_j\,dV_i\,dV_j\,(1 + \xi_{i\,j}),
\end{equation}
where $\xi_{i\,j}$ is the correlation between objects in units $i$ and $j$. 
Thus, equation~(\ref{secondmoment1}) reduces to
\begin{equation}\label{secondmoment2}
\left<N^2\right> = \left<N\right> + {\left<N\right>}^2 + I_2,
\end{equation}
where
\begin{equation}\label{I2}
I_2 = \int\!\!\int n(z_1)\,n(z_2)\,\xi(z_1,z_2,r_{1,2})\,dV(z_1)\,dV(z_2).
\end{equation}
The variance of the count in the box can be expressed as,
\begin{equation}\label{varN}
\frac{\left<N^2\right> - \left<N\right>^2}{\left<N\right>^2} = \frac{1}{\left<N\right>} + \frac{I_2}{\left<N\right>^2} = \sigma_P + \sigma_c,
\end{equation}
can be thought of as the sum of a Poisson part, $\sigma_P$, and a
clustering part, $\sigma_c$.  In the specific case of halos above a given mass threshold, $\xi(z_1,z_2,r_{1,2})$ in equation~(\ref{I2}) is the correlation function between halos with masses (possibly different) above $M_{min}$ at different redshifts, and $n$ is the mass function of such halos.

\section{The Light-Cone Correlation Function}\label{correlation}

While several analytic prescriptions exist for calculating the correlation
function of halos at different redshifts \citep{MW96,Porciani98,SB02}, the
correlation function measured in observations of LBGs is an average along
the light-cone over the survey volume.  The observed correlation between
each halo with mass greater than $M_{min}$ and each other halo with mass
greater than $M_{min}$ is \citep{Matarrese97}:
\begin{eqnarray}\label{xiM}
\xi_{LC,M}^{hh}(>\!\!M_{min},r) \!\!&\!\! = \!\!&\!\! \left<N\right>^{-2} \int\!\!\int dV(z_1)\,dV(z_2)\,\xi^{mm}(\overline{z},r) \nonumber \\ \!\!&\!\! \!\!&\!\! \times n(z_1)\,n(z_2)\,b_{eff}(z_1)\,b_{eff}(z_2),
\end{eqnarray}
where
\begin{equation}\label{beff}
b_{eff}(z) = \frac{\int_{M_{min}}^{\infty} dM\,\frac{dn}{dM}(z, M)\,b(z, M)}{n(z, >\!\!M_{min})}
\end{equation}
is the effective bias used to include halos at all masses above $M_{min}$, $\xi^{mm}$ is the mass autocorrelation function, $n$ is the halo mass function, $b$ is the bias factor, $r$ is the comoving
distance between halos, and $\overline{z} = \overline{z}_M \equiv (z_1 + z_2)/2$.

We use the reformulation of the correlation function on the light-cone given by \citet{YS99} which involves only a single integral:  
\begin{equation}\label{xiYS}
\xi_{LC,YS}^{hh}(>\!\!M_{min},r) = \frac{\int_{z_{min}}^{z_{max}}\,n^2(z)\,\xi^{mm}(z,r)\,dV(z)}{\int_{z_{min}}^{z_{max}}\,n^2(z)\,dV(z)}.
\end{equation}
While the mass function evolves extremely rapidly over the range of the PHSF, the evolution is minimal between two points separated by a distance, $r$, small enough to produce a non-negligible correlation.  This is the key approximation made by \citet{YS99} and hold well even in our regime.

For the correlation function at different redshifts that appears in equation~(\ref{I2}), we use
\begin{eqnarray}\label{xi12}
\xi(z_1,z_2,r_{1,2}) \!\!&\!\! = \!\!&\!\!  \xi^{hh}(>\!\!M_{min},z_1,z_2,r) \nonumber \\
						\!\!&\!\! = \!\!&\!\! \xi^{mm}(r,z=0)\,D(z_1)\,D(z_2) \nonumber \\ 
						\!\!&\!\!  \!\!&\!\! \times b_{eff}(z_2,M_{min})\,b_{eff}(z_1,M_{min}),
\end{eqnarray}
where $D(z)$ is the linear growth factor.  Using instead the measured correlation function along the light-cone given in equation~(\ref{xiYS}) ($\xi(z_1,z_2,r_{1,2}) = \xi_{LC,YS}^{hh}(>\!\!M_{min},r)$), results in a value of $\sigma_c$ different by less than a percent.

\section{A Model for High-Redshift Star-Forming Galaxies}\label{model}

To compare our results for the statistics of halos to those observed, we
need a way of equating the mass of a halo, $M_{halo}$, with the luminosity
of the observable galaxy it contains.  In this section, we review a model
given by \citet{SLE07} that prescribes such
a transformation for LBGs and LAEs.

The SLE07 model associates LBGs and LAEs with merger-activated star
formation in dark-matter halos.  The ratio of baryonic to dark matter mass
in these halos is equal to the cosmic value, $\Omega_b/\Omega_m$.  The
efficiency with which the baryons are converted into stars, denoted by $f$,
is a constant, $f=f_\star$, for halos more massive than a critical value
$M_{halo,crit}$.  However, for halos below this mass, the feedback from
supernovae suppresses star formation such that $f \!=
\!f_{\star}\,(M_{halo}/M_{halo,crit})^{2/3}$.  Modeling and low-redshift observations suggest that $M_{halo,crit}$ corresponds to a velocity in the halo of
$\sim\!100\,{\rm km\,s^{-1}}$ \citep{DW03,Kauffmann03}.  We express the time-scale for star
formation at $z$ as the cosmic time, $t_{H}(z)$, times the star
formation duty cycle, $\epsilon_{DC}$.  $\epsilon_{DC}$ gives
the fraction of the Hubble time during which the star formation occurs.
The average star formation rate is then
\begin{equation}\label{SFR}
\dot{M}_{\star}(M_{halo}) = \frac{f\,(\Omega_b/\Omega_m)\,M_{halo}}{t_H(z)\,\epsilon_{DC}}.
\end{equation}

For LBGs, the luminosity per unit frequency at a wavelength of
$1500\,\mbox{\AA}$ is given by
\begin{equation}\label{LBG}
L_{1500} =
8.0\,\times\,10^{27}\,(\dot{M}_{\star}/M_{\odot}\,
{\rm yr^{-1})\,erg\,s^{-1}\,Hz^{-1}},
\end{equation}
assuming a Salpeter initial mass function (IMF) of stars.

For LAEs with a low metallicity (1/20 solar) and a Salpeter IMF one gets
$N_{ip} = 4 \times 10^{53}$ ionizing photons emitted per $M_{\odot}$ of
star formation per year.  A fraction $1-f_{ip}$ of these photons do not
escape from the galaxy and produce ions, $2/3$ of the resulting
recombinations each produce a Lyman-$\alpha$ photon with energy
$h\,\nu_{Ly\alpha}$, and only a fraction $T_{Ly\alpha}$ these photons
escape into and pass through the intergalactic medium to be observed.  The
Lyman-$\alpha$ luminosity is then
\begin{equation}\label{LAE}
L_{Ly\alpha} = \frac{2}{3}\,h\,\nu_{Ly\alpha}\,T_{Ly\alpha}\,(1 - f_{ip})\,N_{ip}\,\dot{M}_{\star},
\end{equation}
where $\nu_{Ly\alpha}$ is the frequency of the Lyman-$\alpha$ transition.

SLE07 fit the free parameters in their model ($f_{\star}$ and
$\epsilon_{DC}$) to observations at $z\!\sim \!6$.  For LBGs at $z \!\simeq
\!6$, the best fit values and 1-$\sigma$ errors are $f_{\star} =
0.16^{+0.06}_{-0.03}$ and $\epsilon_{DC} = 0.25^{+0.38}_{-0.09}$.  For LAEs
at $z = 6.6$, $f_{\star}\,T_{Ly\alpha} = 0.063^{+0.004}_{-0.018}$ and
$\epsilon_{DC} = 1.0^{+0.0}_{-0.5}$.  These fit parameters are then used to
determine the model at higher redshifts.  We adopt this simple model with a
fixed choice of its free parameters only as an illustrative example for
relating the statistics of dark matter halos to observed galaxies.  All of
the plots given as functions of halo mass in the subsequent sections can be
easily related to galaxy luminosities in the context of any more
complicated models for galaxy formation and evolution.

\section{Results}\label{results}

Next, we present the moments of the true RPD for i-, z-, and J-dropout LGBs
in \S \ref{zdist}.  Having derived expressions for the moments of
halos counts in a survey volume and the correlation function for such halos
along the light-cone in \S \ref{surveycompare}, we may compare these
expressions quantitatively with those derived using a snapshot approach for
various dropout surveys of LBGs in halos of different masses.  The
fractional variation does not depend on the survey field-of-view but only
on the variation along the line-of-sight.  Thus, our results apply to a
wide variety of surveys for LBGs including the Great Observatories Origins
Deep Survey (GOODS), the Hubble Ultra Deep Field (HUDF), and future surveys
using the Subaru Multi-Object Infrared Camera and Spectrograph (MOIRCS) and
the HST Wide Field Camera 3 (WFC3).

For our calculations, we use the mass function given by \citet{ST99} and
the bias factor in \citet{SMT01}.  Our results without the light-cone
effect are produced by assuming that the entire volume exists at the mean
redshift of the PHSF.

\subsection{The Redshift Probability Distribution}\label{zdist}

\begin{figure}
\begin{center}
\includegraphics[width=\columnwidth]{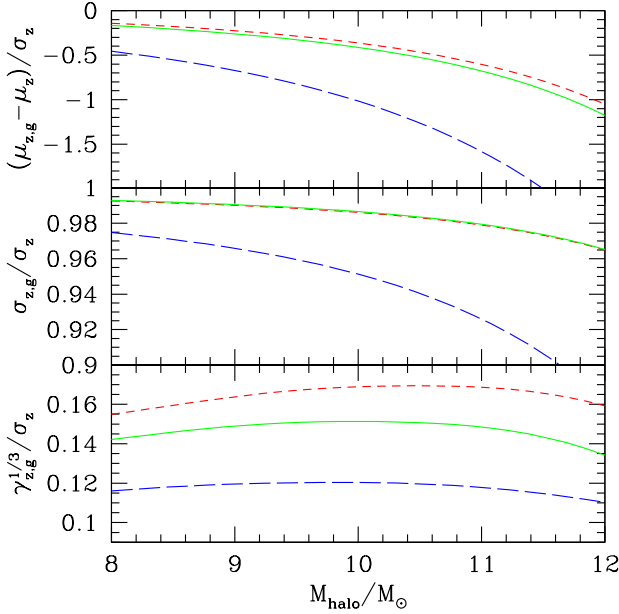}
\caption{Moments of the RPD for the i- (dashed), z- (solid), and J-
(long-dashed) bands normalized by the standard deviation of the PHSF as a
function of host halo mass.  The upper panel shows the difference between
the mean of the RPD and the PHSF, while the center and lower panels plot the
rms variation and asymmetry in the RPD, respectively.}
\label{zdistM} 
\end{center}
\end{figure}

\begin{figure}
\begin{center}
\includegraphics[width=\columnwidth]{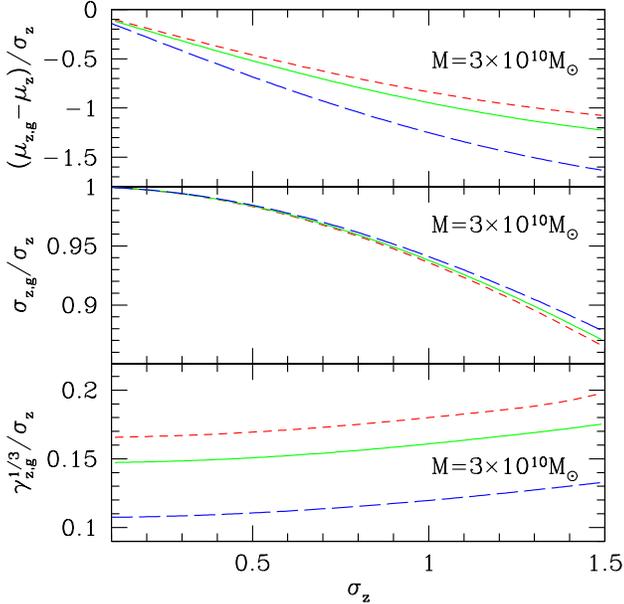}
\caption{Same as figure~\ref{zdistM} except that the moments are plotted as functions of the standard deviation assumed for the PHSF for a fiducial minimum halo mass of $3 \times 10^{10}\,M_{\odot}$.}
\label{zdistS} 
\end{center}
\end{figure}

\begin{figure}
\begin{center}
\includegraphics[width=\columnwidth]{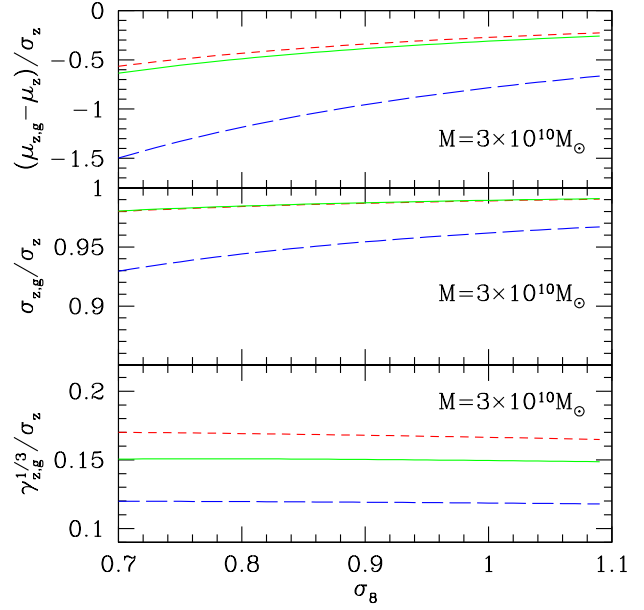}
\caption{Same as figure~\ref{zdistM} except that the moments are plotted as functions of the cosmological parameter $\sigma_8$ for a fiducial minimum halo mass of $3 \times 10^{10}\,M_{\odot}$.}
\label{zdist8} 
\end{center}
\end{figure}

Since the halos that host LBGs are rare at high redshifts and their
abundance varies exponentially with redshift over the width of the PHSFs for
dropout surveys, the RPD of LBGs in such a survey is biased toward lower
redshifts.  The low-redshift tail of a Gaussian PHSF exaggerates this
effect.  The true RPD, $P_g(z)$, is given by equation~(\ref{Pzg}).  The
mean, variance, and skewness of the RPD are given by
\begin{equation}\label{muzg}
\mu_{z,g} = \int_0^\infinity z\,P_g(z)\,dz,
\end{equation}
\begin{equation}\label{sigzg}
\sigma^2_{z,g} = \int_0^\infinity (z-\mu_{z\,g})^2\,P_g(z)\,dz,
\end{equation}
and
\begin{equation}\label{gamzg}
\gamma_{z,g} = \frac{\int_0^\infinity (z-\mu_{z\,g})^3\,P_g(z)\,dz}{\sigma^{3/2}_{z\,g}}. 
\end{equation}
The moments of the RPD for i-, z-, and J-band dropouts are shown in
figures~\ref{zdistM},~\ref{zdistS}, and~\ref{zdist8}.  The plots show their
dependence on halo mass, the broadness of the Gaussian shapes assumed for
the PHSFs, and the cosmological parameter $\sigma_8$.

The most important difference between the PHSF and the RPD for a given band
is in their means.  The exponentially varying mass function biases the
survey toward lower redshifts.  The mean of the RPD is offset from that of
the PHSF by an amount on the order of the PHSF's standard deviation.  Thus,
in a J-dropout survey, the LBGs can be clustered around a redshift of as
low as $z \!\sim \!8.5$, depending on the luminosity of the galaxies
considered, instead of being at $z \!\sim\! 10$.
\begin{figure}
\begin{center}
\includegraphics[width=\columnwidth]{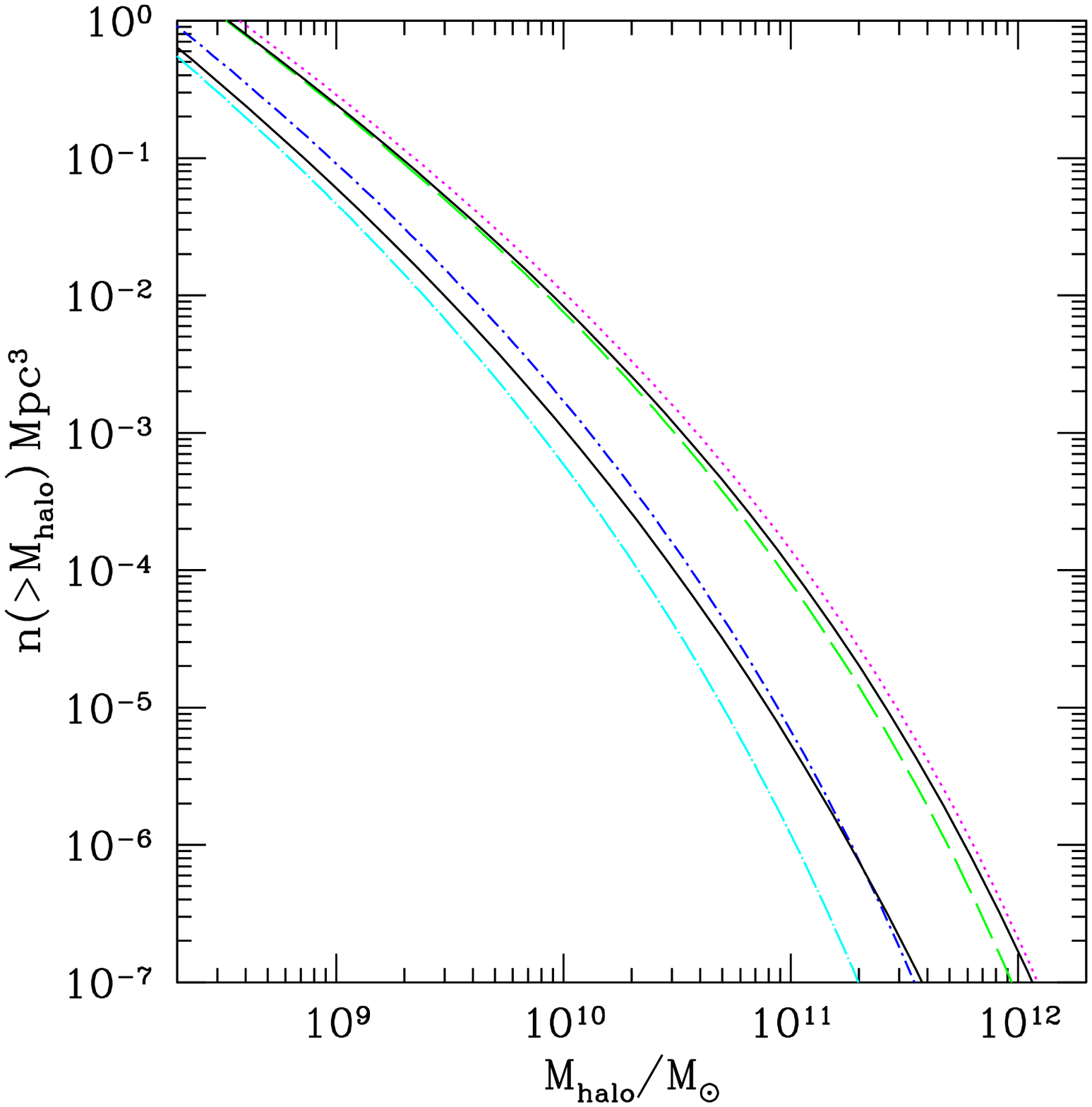}
\end{center}
\caption{\label{mfplot} The mass function of halos extracted from z-
(top solid) and J- (bottom solid) dropout surveys centered at $z\!=\!7.4$ and $z\!=\!10$, respectively.  The underlying Sheth-Tormen mass functions at $z = 7$ (dotted), $7.4$ (long-dashed), $9$ (dot-dashed), and $10$ (dot-long-dashed) are plotted for comparison.}
\end{figure}

Another important point to extract from figure~\ref{zdistM} is that the
LBGs are segregated by their halo mass (luminosity) through the survey.
Since the mean of the RPD decreases monotonically with halo mass
(luminosity), the RPD for more luminous galaxies is shifted to lower
redshifts than that of less luminous ones.  While this series of RPDs
overlaps, it is interesting to note that, for J-dropouts, galaxies in halos
with $M \!\sim\! 10^{8}\,M_{\odot}$ and $M \!\sim\! 10^{11}\,M_{\odot}$ are
almost completely unmixed since the means of their RPDs are about separated
by about one rms variation.  This has important implications for trying to
reconstruct the mass function from observations since different masses are
being observed at different redshifts.  Consider halos in mass bins at two
different masses, $M_{HM}$ and $M_{LM}$, where $M_{HM} \!\!>\!\! M_{LM}$, that are
observed at two different redshifts, $z_{HM}$ and $z_{LM}$, such that
$z_{HM} \!<\! z_{LM}$.  The number of halos with $M=M_{HM}$ is higher than it
would be if these halos were seen at $z_{LM}$.  Thus, the resulting
inferred mass function is shallower than the true one determined using
halos that are all at the same redshift.  The effect is lessened by
supression due to the PHSF, however.  The density of halos of a given mass
is less than the underlying density of such halos where they are seen at
$\mu_{z, g}$.  Yet, their density is still higher than the underlying
density at $\mu_z$.  The amplitudes of the mass function and the RPD at
various redshifts can be compared for J-band halos with $M_{halo} > 2
\times 10^{12} M_{\odot}$ in figure~\ref{Nplot}.  Figure~\ref{mfplot}
compares the extracted mass function with the underlying Sheth-Tormen mass
function used to compute it.

The RPDs are slightly (by $\la 10$ percent) narrower than the PHSFs.  The
normalized rms variation in the RPD as a function of the standard deviation
of the PHSF is nearly identical for each band as are the equivalent plots
as a function of halo mass for the i- and z- bands.  This indicates that
the slight narrowing is independent of the assumed standard deviation of
the Gaussian PHSFs and depends only on the target redshift of a given band,
i.e. the mean of its PHSF.

The asymmetry of the RPD is plotted in Figures~\ref{zdistM},~\ref{zdistS}
and~\ref{zdist8} as $\gamma_{z,g}^{1/3}/\sigma_z$.  The asymmetry is very
small, on the order of tens of percent.  This degree of symmetry and the
fact that the skewness is positive might seem counter-intuitive since the
exponentially varying mass function should bias the RPD toward lower
redshifts.  This bias, however, is manifested in the shift in the mean of
the RPD away from that of the PHSF rather than in skewing the RPD.  The
Press-Schechter mass function with a simplified growth factor,
$D(z)=1/(1+z)$, has two dependencies on redshift, a linearly increasing
factor and an exponentially decaying one that dominates at high-redshift,
$n = A\,(1+z)\,e^{-B\,(1+z)^2}$, where $A$ and $B$ are independent of
redshift.  The exponential decay factor, however, is simply the tail of a
Gaussian, which when multiplied by the Gaussian PHSF yields another
symmetric Gaussian.  Only the subordinate linear factor contributes to the
asymmetry.

Finally, although changing the value of $\sigma_8$ has a small effect on
the value of the parameters we calculated, the difference as shown in
Figure~\ref{zdist8} was not drastic; all of the basic results we have just
presented remain unchanged.  In fact, the asymmetry of the RPD is virtually
unchanged when $\sigma_8$ is varied.  This is consistent with the asymmetry
being due only to the linear dependence on redshift, as discussed above,
and not to the exponentially varying factor, which contains most of the
mass function's dependence on $\sigma_8$.

\subsection{The Light-Cone Effect on Moments of Counts 
and the Correlation Function}\label{surveycompare}

\begin{figure}
\begin{center}
\includegraphics[width=\columnwidth]{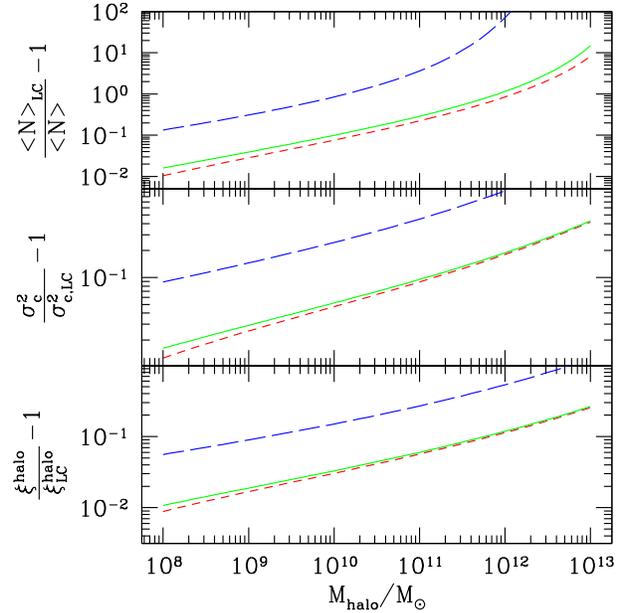}
\end{center}
\caption{\label{LCplot} The fractional effect of the light-cone on the mean
number, clustering variance, and correlation function of halos in a dropout
survey for LBGs in the i- (dashed), z- (solid), and J- (long-dashed) bands.
The clustering variance plotted here is $\sigma_{c} = I_2/\left<N\right>^2$.}
\end{figure}

As described above, the equations for the moments of the count of objects
in the survey and their correlation are different if the light-cone effect
is included.  This effect is greatly enhanced by the wide PHSFs of the bands
used in dropout surveys.  For these surveys, the volume element in
equations~(\ref{Nave}),(\ref{I2}), and~(\ref{xiYS}) is replaced via
\begin{equation}\label{dVsub}
dV(z) \rightarrow \frac{e^{-\frac{(z-\mu_z)^2}{2\,\sigma_z^2}}}{\sqrt{2\,\pi\,\sigma_z^2}}\,\frac{dV(z)}{dz}\,dz.
\end{equation}
Figure~\ref{LCplot} shows the results of including this effect on the mean
count, clustering variance, and correlation function of halos containing
the LGBs in i-, z-, and J-band dropout surveys.  While the value of each of
these statistics varies depending on the field-of-view of the particular
survey under consideration, the fractional effect is independent of
field-of-view since a change in the area of the survey is orthogonal to the
variation in the mass and correlation functions along the lightcone.

\begin{figure}
\begin{center}
\includegraphics[width=\columnwidth]{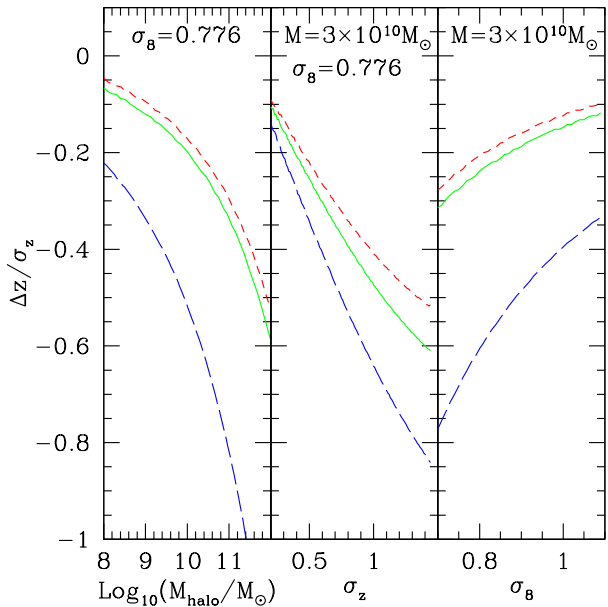}
\end{center}
\caption{\label{dzplot} The difference between the redshift at which the
underlying mass function gives the same density of halos as observed in i-
(dashed), z- (solid), and J- (long-dashed) band dropout surveys and the
mean of the PHSF.  The left panel plots the difference as a function of
minimum halo mass, while the center and right panels show it as functions
of the assumed PHSF standard deviation and $\sigma_8$, respectively, for a
fiducial minimum halo mass of $3 \times 10^{10}\,M_{\odot}$.}
\end{figure}

For i- and z-dropouts, the light-cone effect can make an order unity difference in the mean number of objects and a difference of roughly ten percent in the clustering variance of the count and the correlation
(Eq.~\ref{xiYS}) at the high mass end.  The effect for J-dropouts is much larger (by almost an order-of-magnitude) because the mass function has more evolution over the J-band both due to the increased steepness of the mass function at higher redshift and because the PHSF is much broader for the the J-band.  The variation in the effect with halo mass is due to the mass segregation discussed in the previous section.  In particular, the mass dependence of the effect on $\left<N\right>$ is manifested in the flattening of the mass function extracted from the survey shown in Figure~\ref{mfplot}.

In associating our calculations for the halo correlation function with
LBGs, we note that the halos hosting LBGs do not constitute a fair sample
of the entire halo population.  \citet{ST03} show through numerical
simulations that these halos have undergone substantial accretion in their
recent past giving them an extra ``temporal" bias.  While these simulations
were performed at $z=3$, there is as of yet no analytical method for
predicting this extra bias at higher redshift.  Since we compute only the
fractional difference in the variance and correlation function, we are safe
in ignoring this effect as long as the bias of LBGs over halos does not
vary much within the redshift range of the RPD.

Finally, in \S \ref{zdist} we showed that the LBGs in dropout surveys,
which are more numerous than would be calculated ignoring the light-cone
effect, are distributed at lower redshifts than indicated by the PHSF.
However, it is also interesting to ask at what redshift does the 
``snapshot" calculation yield the same number of LBGs as the light-cone
calculation.  Hypothesizing a narrow-band survey for LBGs, this is
equivalent to asking at what redshift does this hypothetical survey give
the same density of LGBs as a dropout survey.  The difference between this
redshift and the mean of the PHSF is plotted in Figure~\ref{dzplot} for
each band.

\section{Discussion}\label{discussion}

The variation of the mass function of halos along the light-cone within the
volume of a dropout survey results in a mass segregation effect that is also manifested in their hosted LBGs.  The
observed sources are not at the mean of the photometric selection function
(PHSF) but instead are distributed at lower redshifts (see
Fig. \ref{Nplot}).  This effect applies to true LBG dropouts and ignores possible interlopers
from lower redshifts whose spectra mimic those of higher redshift LGBs.
The mass segregation results in different measured statistics of LBGs from
those expected from theory or simulations along a space-like slice
(snapshot) through the universe at the ``mean survey redshift."  In
particular, the mass/luminosity function extracted from such a survey is
shallower than the underlying mass/luminosity function because of the
different strengths of the light-cone effect on halos of different masses (LBGs of different
luminosities).  This flattening of the mass function is particularly important
in the J-band because of its larger width in redshift space (see
Fig. \ref{mfplot}).

Without spectroscopic measurements to confirm the redshifts of a sample of
LBGs, the mass segregation effect presents an added complication in
continuing efforts to determine the effect of high-redshift LBGs on
reionization \citep{Nagamine06,SLE07} and the microwave background
\citep{BL07}, or efforts to measure the high-redshift evolution of the
star-formation rate \citep{ST06,Ellis07}.  Reionization is a highly
inhomogeneous process, and so the light-cone effect on the correlation
function is also particularly relevant in that context \citep{BL04,FL05}.  The
lightcone effect on the measured correlation function of LBGs would also be
important in attempts to use the clustering properties of LBGs to infer the
masses of their host halos at higher redshift.

\citet{Dow-Hygelund07} made an effort to follow up spectroscopically on
LBGs from i-band dropout surveys near $z\! \simeq \!6$ to measure their
redshifts precisely.  However, they were only able to confirm redshifts on
6 LBGs in their sample, a number insufficient to trace the RPD in a
statistically significant way.  \citet{Ando04,Ando07} preform similar
studies on LBGs near $z \!\simeq\! 5$, but the sample size they obtained
was also not sufficiently large for this purpose.  Ideally, one would like
to perform this type of analysis on z- or J-band dropouts with the goal of
tracing the RPD, but this would be very difficult both because of the
present lack of candidates and because of the high integration times
necessary with current technology.

\section{Acknowledgements}

We would like to thank Rychard Bouwens and Dan Stark for useful
discussions.  JM acknowledges support from a National Science Foundation
Graduate Research Fellowship. This research was supported in part by
Harvard University funds.

\end{document}